\documentclass[aps,prb,superscriptaddress,showpacs,twocolumn]{revtex4}

\usepackage{amsmath,amsfonts,amssymb,graphics,graphicx,epsfig,color,times,indentfirst,layout}

\begin{document}

\title{Implementing Genuine Multi-Qubit Entanglement of Two-Level-System Inside a Superconducting Phase Qubit}

\author{Long-Bao Yu}

\affiliation{Laboratory of Quantum Information Technology, ICMP
and SPTE, South China Normal University, Guangzhou, China}

\author{Zheng-Yuan Xue}

\affiliation{Department of Physics and Center of Theoretical and
Computational Physics, The University of Hong Kong, Pokfulam Road,
Hong Kong, China}

\author{Z. D. Wang}

\affiliation{Department of Physics and Center of Theoretical and
Computational Physics, The University of Hong Kong, Pokfulam Road,
Hong Kong, China}

\author{Yang Yu}
\affiliation{National Laboratory of Solid State Microstructures
and Department of Physics, Nanjing University, Nanjing, China}

\author{Shi-Liang Zhu}
\email{slzhu@scnu.edu.cn} \affiliation{Laboratory of Quantum
Information Technology, ICMP and SPTE, South China Normal
University, Guangzhou, China}

\begin{abstract}
The interaction between a superconducting phase qubit and the
two-level systems locating inside the Josephson tunnel barrier is
shown to be described by the XY model, which is naturally used to
implement the iSWAP gate. With this gate, we propose a scheme to
efficiently generate genuine multi-qubit entangled states of such
two-level systems, including multipartite W state and cluster
states. In particularly, we show that, with the help of the phase
qubit, the entanglement witness can be used to efficiently detect
the produced genuine multi-qubit entangled states. Furthermore, we
analyze that the proposed approach for generating multi-qubit
entangled states can be used in a wide class of candidates for
quantum computation.

\end{abstract}


\date{\today}

\maketitle 

\section{Introduction}

Generation of entangled states of an increasing number of qubits has
been an important goal and benchmark in the field of quantum
information\cite{Blatt1,nc}. Multi-qubit entangled states serve as
the essential physical resources for measurement-based quantum
computing \cite{Raussendorf,Briegel} and quantum error-correcting
codes\cite{Shor,Steane}. Some of them, 
such as W state \cite{Dur}, GHZ state \cite{Greenberger} and cluster
state\cite{Raussendorf,Briegel}, have been investigated both
theoretically and
experimentally~\cite{Haffner,Lu,Walther,Roos,Zhu2005}; however, the
experimental preparation of multi-particle entanglement has been
proved to be extremely challenging. To date, the entangled states up
to eight atoms \cite{Haffner} or six photonic qubits\cite{Lu} have
been experimentally reported. As for solid-state systems, due to the
difficulty to decouple the qubits with the environments, only the
two-qubit entanglement of the supercoducting qubits has been
demonstrated in experiments\cite{Steffen,McDermott,Berkley}.
Therefore, generation of up to ten qubits entangled states of the
candidates for solid-state quantum computation will be a next
significant and very challenging step towards quantum information
processing.

Among the solid-state systems, superconducting circuit is one of
the most promising candidates  served as hardware implementation
of quantum computers~\cite{Makhlin,You,Yu}. But the short
coherence time limits both of the qubit state manipulation and
information storage. The loss of quantum coherence in the most of
solid state qubits is mainly due to the unwanted coupling of the
qubits with the environments. In particular, the coherence time
would be decay
quickly with increasing number of qubits because each qubit
usually has a control and a measurement circuit. Besides
fulfilling the manipulations and measurements required for the
necessary information processing, all of circuits can also disturb
the qubits and lead to decoherence. Therefore, a possible way to
experimentally prepare more than two-qubit entangled states of
solid state systems, which have not yet been demonstrated, may
need to suppress the decoherence from the environments, such as to
reduce the number of control and measurement lines.

In this paper, we propose a distinct scheme to prepare genuine
multipartite entangled states for several to ten qubits, while the
decoherence from the control and measurement lines may be minimized
by using single control and measurement setup. The system we have in
mind is several to ten of two-level systems
(TLSs)\cite{Simmonds,Cooper,Martinis,Yu_Zhu,Neeley,Martin,Tian,Steffen2,Martinis2,Zagoskin}
locating inside a superconducting phase qubit, e. g., a
current-biased Josephson junction (CBJJ)\cite{Makhlin,You,Yu}.
Recent experiments \cite{Simmonds,Cooper,Martinis,Yu_Zhu,Neeley}
have shown that some of TLSs locate inside the Josephson tunnel
barrier, while the parameters for such TLSs can be detected through
spectroscopic measurements. The lifetime of the TLS is much longer
than the decoherence time of the phase qubit, thus TLS can be used
as high quality quantum memory \cite{Neeley}. Furthermore,
macroscopic quantum jump\cite{Dehmelt,Blatt,Plenio} has been
experimentally demonstrated   for a hybrid model consisting of a
phase qubit and a TLS inside the Josephson tunnel
barrier\cite{Yu_Zhu}. In particular, it has been proposed that the
TLS itself can be used as qubits for quantum
computation\cite{Zagoskin}, and typical operations required for the
information processing, such as the state initialization, universal
logical gate operations and readout, have been experimentally
demonstrated\cite{Neeley}.

Motivated by the progress, in this paper, we show that the
interaction between the superconducting phase qubit and the
two-level systems may be described by the XY model, which is
naturally used to implement the iSWAP gate. With this gate, we can
effectively generate virous genuine multi-qubit entangled states for
TLSs, including the $W$ state and multipartite cluster states.
Moreover, the states of TLSs can be manipulated by controlling the
interaction between them and CBJJ. Such CBJJ-TLS coupling system
offers a natural candidate to realize quantum information processing
through combining the advantages of microscopic and macroscopic
scale systems. In particularly, we found that such entangled states
may be efficiently detected by measuring the entanglement
witness\cite{Toth,Guhne,Chen} with the help of the phase qubit.
Finally, the proposed approach to produce genuine multi-qubit
entanglement may be applied to various systems for quantum
computing, including the trapped ions\cite{Cirac} and
$C_{60}$\cite{Benjamin} etc..

The paper is organized as follows. In sec. II, we briefly introduce
the superconducting phase qubit and TLSs locating inside the
Josephson tunnel barrier, and then we show that such TLSs can serve
as qubits for information processing. Furthermore, we demonstrate
that an iSWAP gate between the phase qubit and each of TLSs can be
achieved. In Sec. III, a scheme to implement genuine multi-qubit
entanglement of such TLSs is proposed. In Sec.IV, the detection of
the prepared multi-qubit entangled states based on the entanglement
witness is studied. In Sec. V, we show that the proposed approach
can be used to generate genuine multi-qubit entangled states in a
wide class of quantum systems, and the paper ends with a brief
discussion.

\section{Two-level-systems inside a superconducting phase qubit}

The system we consider is a hybrid consisting of a standard
superconducting phase qubit and several to ten TLSs inside the
Josephson tunnel barriers, as shown in Fig.1. The superconducting
phase qubit is a CBJJ, and recent experiments have shown that some
of TLSs are located inside the Josephson tunnel barriers.
Furthermore, such TLSs can be considered as qubits for the
information processing, whereas the Josephson phase qubit itself is
a 'register' qubit capable of general logic operations between TLSs
qubits\cite{Neeley}. The Hamiltonian of the phase qubit as shown in
Fig.1a reads
$$H_p=\frac{1}{2C}\hat{Q}^{2}-\frac{I_{0}\Phi_{0}}{2\pi}\cos\hat{\delta}-\frac{I\Phi_{0}}{2\pi}\hat{\delta},$$
where $I_{0}$ is the critical current of the Josephson junction,
$I$ is the bias current, $C$ is the junction capacitance,
$\Phi_{0}=h/2e$ is the flux quantum, $\hat{Q}$ and $\hat{\delta}$
are the charge and gauge-invariant phase difference across the
junction, which obeys the convectional quantizing commutation
relation $[\hat{\delta},\hat{Q}]=2ei$. For large area junctions,
the Josephson coupling ennergy $E_{J}=I_{0}\Phi_{0}/2\pi$ is much
larger than the single charging energy $E_{C}=e^{2}/2C$. The phase
is a well defined macroscopic variable and quantum behavior can be
observed when the bias current is slightly smaller than the
critical current. In this regime, the two lowest energy levels,
$|0\rangle$ and $|1\rangle$, are usually employed as two quantum
states to form a so-called phase qubit. Truncating the full
Hilbert space of the junction to the qubit subspace, the phase
qubit Hamiltonian can be written as,
\begin{equation}\label{1}
H_{P}=-\frac{1}{2}\omega_{10}\sigma_{z}.
\end{equation}
where $\omega_{10}$ is the frequency difference between
$|0\rangle$ and $|1\rangle$. The states of the qubit can be fully
controlled with the bias current in the form of
$$I(t)=I_{dc}+I_{lf}(t)+I _{\mu w c}(t)\cos\omega_{10}t+I _{\mu
ws}(t)\sin\omega_{10}t,$$ where  the classical bias current is
parameterized by  $I _{\mu wc}$, $I _{\mu ws}$ and $I_{dc}$.

\begin{figure}
\includegraphics[width=6.5cm, angle=90]{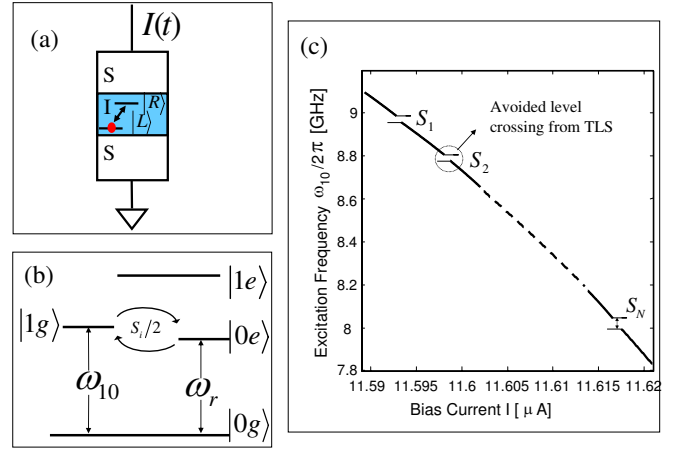}
\caption{(Color online) The phase qubit and TLSs locating inside the
Josephson junction tunnel barrier. (a) Schematic show of the hybrid
system. The system is a CBJJ coupled with some embedded TLSs. (b)
Schematic energy level diagram for a junction coupled to a TLS. The
ground state (the excited state) of the CBJJ is denoted as
$|0\rangle$ ($|1\rangle$), and the frequency difference is
$\omega_{10}$; $|g\rangle$ and $|e\rangle$ represent the ground
state and the excited state of TLS with level spacing $\omega_{r}$.
(c)Schematic show that Qubit frequency $\omega_{10}/2\pi$ vs current
bias for the capacitively shunted design tunnel junction. Similar to
the experimental observation, we plotted about ten splittings
observed in the spectroscopy.} \label{FIG.1}
\end{figure}

The transition frequency $\omega_{01}$ can be measured using
spectroscopy \cite{Neeley} and is a continuous function of bias
current. It was found that some TLSs may locate inside the Josephson
tunnel barrier (see Fig.1(a)). A TLS is understood to be an atom, or
a small group of atoms, that tunnels between two lattice
configurations\cite{Phillips,Neeley}. A TLS can induce an energy
splitting in the curve of the function $\omega_{01}(I)$ , as shown
in Fig. 1(c). The splittings $\Delta_j \in [20, 100]$ MHz were
observed and they are separated by $\delta f\sim 200$ MHz on average
\cite{Martinis} in the spectroscopy. Therefore, one can characterize
the positions and sizes of TLS in the CBJJ energy spectrum through
spectroscopic measurements. When the register qubit is detuned from
the TLS by $\delta f$, the effective coupling strength is described
by $\Delta_j^{2}/4\delta f$. Therefore, with a splitting magnitude
of several tens of MHz, it is reasonable to assume that only a
single TLS satisfies the near-resonance condition while the other
TLSs are far off-resonance. Furthermore, as the number of TLS mainly
depend on the tunnel junction area, the design with small area and
external low-loss capacitor (keeping the critical current constant)
will greatly improve its performance \cite{Steffen2}. It is now
possible to obtain a tunnel junction within about ten useful TLSs in
the CBJJ spectroscopy ranging over $\sim 2$ GHz with the improved
design \cite{Steffen2}.

 As shown in Fig. 1(a), each TLS can be modeled as a charged particle that
can tunnel between two nearby different positions with different
wave functions $|R\rangle$ and $|L\rangle$ (correspond to critical
currents $I_{c}^{R}$ and $I_{c}^{L}$) within the tunnel barrier. The
interaction Hamiltonian between the resonators and the
critical-current is \cite{Simmonds}
\begin{equation}\label{2}
H_{int}=-\frac{I_{c}^{R}\phi_{0}}{2\pi}\cos\hat{\delta}\otimes
|R\rangle\langle R|
-\frac{I_{c}^{L}\phi_{0}}{2\pi}\cos\hat{\delta}\otimes|L\rangle\langle
L|.
\end{equation}
Assume a symmetric potential with energy separated by
$\hbar\omega_{r}^{i}$ for the $i$th TLS, then the ground and
excited states are $|g\rangle\simeq(|R\rangle+|L\rangle)/\sqrt{2}$
and $|e\rangle\simeq(|R\rangle-|L\rangle)/\sqrt{2}$. In practical
experiments, the junction is biased near its critical current,
$\delta\rightarrow\delta+\pi/2$ with $\delta\ll1$, thus
$\cos\hat{\delta}\rightarrow\hat{\delta}$. In the basis
$\{|0g\rangle,|1g\rangle,|0e\rangle,|1e\rangle\}$, as shown in
Fig.1(b), denoting $\delta_{ij}=\langle i|\hat{\delta}|j\rangle$,
usually $\delta_{ii}\approx0$ and
$\delta_{01}=\delta_{10}=\frac{2\pi}{\Phi_{0}}\sqrt{\frac{\hbar}{2\omega_{10}C}}$.
Then the Hamiltonian of CBJJ-TLS system becomes\cite{Yu_Zhu}
\begin{equation}\label{4}
H_{total}=-\frac{\hbar\omega_{10}}{2}\sigma_{z}-\sum_{j}\left[\frac{\hbar\omega_{r}^{j}}{2}\tilde{\sigma}_{z}^{j}
+S_{j}\sigma_{x}\tilde{\sigma}_{x}^{j}\right],
\end{equation}
where the effective coupling strength between the CBJJ and the
$j$th TLS is $S_{j}=
\frac{(I_{c}^{R}-I_{c}^{L})}{2}\sqrt{\frac{\hbar}{2\omega_{10}C}}$.
The Pauli matrices $\tilde{\sigma}_{x,z}^j$ operate on the $j$th
TLS states. The CBJJ and the $j$th TLS are tuned into resonance
when the coupling term satisfy
$|\hbar\omega_{10}-\hbar\omega_{r}^{j}|<S_{j}$. The parameters
 $\omega_{r}^{j}$ and $S_{j}$ of the $j$th TLS are usually unknown,
but they can be independently determined by the energy splitting
\cite{Simmonds,Yu_Zhu} on the spectroscopy, as shown in Fig.1.

In the interaction picture and under the rotating-wave
approximation, the effective interaction between the $j$th TLS and
the phase qubit is given by
\begin{equation}\label{4}
\hat{H}_{int}^{j}=-\frac{S_{j}}{2}(\sigma_{x}\tilde{\sigma}_{x}^{j}+\sigma_{y}\tilde{\sigma}_{y}^{j}).
\end{equation}
So the effective CBJJ-TLS interaction is described as the XY
model. Assume that the interaction between CBJJ and the $j$th TLS
is tuned to the resonance with time $t$, one obtain the evolution
operator given by
\begin{equation}
\label{U_gate} U_j(t)=\exp\left[\frac{it S_j}{2}
(\sigma_{x}\tilde{\sigma}_{x}^{j}+\sigma_{y}\tilde{\sigma}_{y}^{j})\right]
\end{equation}
between them. This operator results in an oscillation between
$|1g\rangle$ and $|0e\rangle$ at a frequency $S_j$, \emph{i.e.},
\begin{eqnarray}\label{U_operator}
&&|0g\rangle\rightarrow|0g\rangle, |1e\rangle\rightarrow|1e\rangle,\nonumber\\
&&|1g\rangle\rightarrow\cos(S_jt)|1g\rangle-i\sin(S_jt)|0e\rangle,\nonumber\\
&&|0e\rangle\rightarrow\cos(S_jt)|0e\rangle-i\sin(S_jt)|1g\rangle.
\end{eqnarray}
Note that the iSWAP gate between CBJJ and TLS can be obtained when
$t_j=\pi/(2S_j)\equiv\tau_j$. In the paper, the TLS states,
$|g\rangle$ and $|e\rangle$, represent our logic qubit. The
experiments have shown that the lifetime of such qubits is
sufficiently long to carry out precise gate operations between the
phase qubit and TLSs. In addition, we assume that the states of
different TLSs are well separated from each other in frequency.
Therefore, by adjusting the bias currents, the phase qubit and a TLS
can be tuned into and out of resonance (turning on and off their
coupling), i.e., it allows the independent manipulation of each TLS.

\section{Implementing the $W$ and cluster states of two-levle-systems }

 We now turn to demonstrate that the genuine entanglement of several to ten TLSs
could be generated through the unitary operator described in
Eq.(\ref{U_operator}) by controlling the interaction time between
CBJJ and TLSs.

We assume that there are $N$ TLSs locating inside the Josephson
tunnel barrier, while the coupling constant $S_j$
$(j=1,2\cdots,N)$ have been measured by using the spectroscopy. We
first show that the N-qubit W state can be achieved simply by two
steps.

The first step is to initialize each TLS qubit in any general
state $|\psi\rangle_j=\alpha_j|g\rangle_{j}+\beta_j|e\rangle_{j}$
to the ground state $|g\rangle_j$ by performing an iSWAP operation
between this TLS and the phase qubit. We have demonstrated in the
previous section that such iSWAP gate can be realized by switching
on the interaction between the $j$th TLS and the phase qubit with
the fixed time $\tau_j\equiv \pi/2S_j$. The ground state of the
$j$th TLS is initialized by the following transformation
\begin{equation}\label{I_state}
[iSWAP]_{(j,P)}|\psi\rangle_j \otimes |0\rangle \rightarrow
|g\rangle_{j}\otimes (\alpha_j |0\rangle-i\beta_j|1\rangle)
\end{equation}
with $[iSWAP]_{(j,P)}$ denoting the iSWAP operation between the
$j$th TLS and the phase qubit, provided that the initial state of
the phase qubit is $|0\rangle$. The ground state
$\bigotimes_{j=1}^{N}|g\rangle_{j}$ for all TLSs is then realized
after an iSWAP gate is performed between each TLS and the phase
qubit, hybrid an operation to initialize the phase qubit to the
state $|0\rangle$ between two iSWAP gates.

In the second step, the phase qubit is first tuned far
off-resonance with all TLSs and excited on the $|1\rangle$ state.
Then
 the phase qubit is adiabatically tuned into resonance with one of
TLSs (f.g., the $j$th TLS), effectively turning on the coupling
between this TLS and the phase qubit with the time $t_{j}$.
Sequentially, the register qubit is interacted with every TLSs
only once with appropriate time $t_j$, and then an operator
$U_{j}=e^{-iH_{int}^j t_{j}/\hbar}$ between the phase qubit and
the $j$th TLS is performed. In this case (as shown in the
Fig.2(a)), the final state of the TLSs and the phase qubit becomes
\begin{eqnarray}
&&\prod_{l=1}^{N}U_{l}|1\rangle\bigotimes_{j=1}^{N}|g\rangle_{j}\rightarrow\nonumber\\
&&\prod_{l=2}^{N}U_{l}(\cos(S_{1}t_{1})|1\rangle|g\rangle_{1}-
i\sin(S_{1}t_{1})|0\rangle|e\rangle_{1})\bigotimes_{j=2}^{N}|g\rangle_{j}\nonumber\\
&&\rightarrow\prod_{l=3}^{N}U_{l}[\cos(S_{1}t_{1})|g\rangle_{1}(\cos(S_{2}t_{2})|1\rangle|g\rangle_{2}\nonumber\\
&&- i\sin(S_{2}t_{2})|0\rangle|e\rangle_{2})-
i\sin(S_{1}t_{1})|e\rangle_{1}|g\rangle_{2}|0\rangle]\bigotimes_{j=3}^{N}|g\rangle_{j}\nonumber\\
&&\ \ \ \ \ \ \ \ \ \ \ \ \ \ \ \ \ \ \ \ \ \ \ \ \ \vdots\nonumber\\
&&\rightarrow-i|0\rangle
\sum_{l=1}^{N}\prod_{j=1}^{l-1}\cos(S_{j}t_{j})\sin(S_{l}t_{l})\bigotimes_{k\neq
l}^{N}|g\rangle_{k}|e\rangle_{l}\label{W_state}.
\end{eqnarray}
If the interaction time between the $j$th TLS and the phase qubit
is chosen specifically as
$t_{j}=\frac{1}{S_{j}}\arcsin(\frac{1}{\sqrt{N+1-j}})$, i.e.,
$\prod_{j=1}^{l-1}\cos(S_{j}t_{j})\sin(S_{l}t_{l})=\frac{1}{\sqrt{N}}$,
then the final state of Eq.(\ref{W_state}) indeed becomes the
standard W-state of the $N$ TLSs described by
$|W\rangle_{N}=\frac{1}{\sqrt{N}}\sum_{l=1}^{N}\bigotimes_{k\neq
l}^{N}|g\rangle_{k}|e\rangle_{l}$.

By using the above process, two important entangled states may be
realized. Firstly, the Bell state of two arbitrary TLSs $j$ and
$k$ can be achieved as
\begin{eqnarray}\label{Bell}
U_j U_k |1\rangle |g\rangle_{j} |g\rangle_{k
}\rightarrow&\frac{-i}{\sqrt{2}}(|g\rangle_{j}|e\rangle_{k}+
|e\rangle_{j}|g\rangle_{k}) |0\rangle,
\end{eqnarray}
when the interaction time are chosen as $t_{j}=\tau_{j}/2$ and
$t_{k}=\tau_{k}$.  Secondly, the W state of arbitrary three TLSs,
such as $j$, $k$ and $l$-th TLS, may also be obtained through
\begin{eqnarray}
&&U_j U_k U_l |1\rangle |g\rangle_{j} |g\rangle_{k} |g\rangle_{l}\rightarrow\nonumber\\
&&\frac{-i}{\sqrt{3}}(|g\rangle_{j}|g\rangle_{k}|e\rangle_{l}+
|g\rangle_{j}|e\rangle_{k}|g\rangle_{l}+
|e\rangle_{j}|g\rangle_{k}|g\rangle_{l}) |0\rangle
\end{eqnarray}
by choosing the interaction time $t_{j}=\tau_{j}/3$,
$t_{k}=\tau_{k}/2$, and $t_{l}=\tau_{l}/2$, respectively.

Furthermore, the cluster state
$|C_{N}\rangle\equiv\frac{1}{2^{N/2}}\bigotimes_{j=1}^{N}(|g\rangle_{j}\sigma_{z}^{j+1}+|e\rangle_{j})$
of $N$ TLSs may be implemented by a similar process. The process
is also two steps. (i) The first step is to initialize the $N$th
TLS to the ground state $|g\rangle_{N}$ and all the other TLSs to
the states
$|+\rangle_j=\frac{1}{\sqrt{2}}(|g\rangle_j+|e\rangle_j)$, while
initialize the phase qubit to
$|+\rangle_P=\frac{1}{\sqrt{2}}(|0\rangle+|1\rangle)$. (ii)
Sequentially perform the iSWAP operations between CBJJ and each of
TLSs. Note that, to cancel the single-qubit phase factor, a
$Z^{\frac{\pi}{2}}_P$ ( which denotes a $\pi/2$ rotation around
the $z$ axis) pulse needs to be applied to the phase qubit both
before and after each iSWAP operation. After that, all $N$ TLSs
may be connected to make a large cluster chain (as shown in the
Fig.2(b)), i.e.,
\begin{eqnarray}\label{}
&&[iswap]_{(N,P)}\prod_{j=1}^{N-1}\{Z_{P}^{\frac{\pi}{2}}[iswap]_{(j,P)}[Z]_{P}^{\frac{\pi}{2}}\}\nonumber\\
&&\{\bigotimes_{j=1}^{N-1}|+\rangle_{j}\}\otimes|g\rangle_{N} |0\rangle\nonumber\\
&&=\frac{1}{2^{N/2}}\bigotimes_{j=1}^{N}(|g\rangle_{j}\sigma_{z}^{j+1}+|e\rangle_{j}) |0\rangle\nonumber\\
&&=|C_{N}\rangle |0\rangle.
\end{eqnarray}

Moreover, The chain cluster states can be connected to produce
higher dimensions cluster states by repeating iSWAPs
\cite{Tanamoto}.

\begin{figure}
\includegraphics[scale=0.35, angle=90]{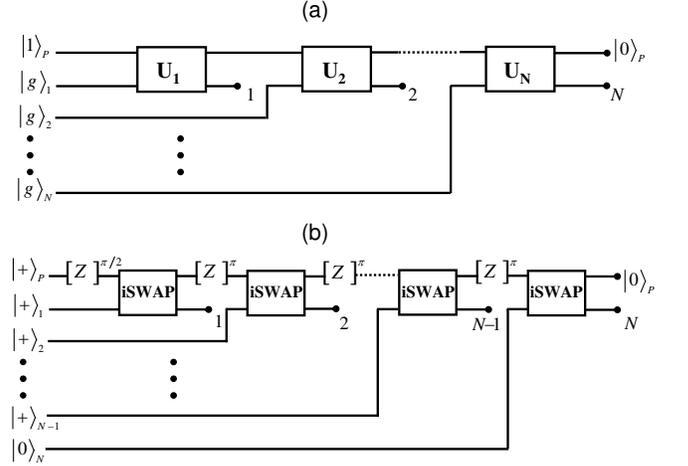}
\caption{Quantum circuits for generating (a) W state and (b)
cluster state of N-TLSs.} \label{FIG.2}
\end{figure}

\section{Detection of the genuine multi-qubit enetanglement }

To ensure the obtained state is the desired multi-qubit entangled
state, one must detect the state of TLSs. Because the TLSs cannot
be directly measured, one must previously perform iSWAP operation
to transform the state of the $j$th TLS to the phase qubit (up to
a correctable Z rotation), then the information can be read out
through measuring the phase qubit with quantum state tomography
(QST) \cite{Neeley}. Currently the measurement \cite{Cooper} of
the phase qubit  with high-fidelity ($F=0.96$) can be completed
with a short time (less than $5$ ns). The readout technique is
achieved by applying a short bias current pulse $\delta I(t)$ that
adiabatically reduces the well depth $\Delta U/\hbar\omega_{p}$,
so that the first excited state lies very near the top of the well
when the current pulse is at its maximum value. In this way, one
can read out the states of TLSs one by one.

 For detecting
two-qubit entangled states, one can use tomographic state analysis
\cite{Steffen,Roos,Berkley} to reconstruct the density matrix with
near-unity detection efficiency. This is achieved by single qubit
rotations and subsequent projective measurements. For the
two-qubit system, a convenient set of operators is given by 16
operators $\sigma^{(1)}_{\alpha}\otimes\sigma_{\beta}^{(2)}$
($\alpha,\beta=0,x,y,z$), where $\sigma^{(j)}_{\alpha}$ denote
Pauli matrices of qubit $j$. The reconstruction of the density
matrix $\rho$ is accomplished by measuring the expectation values
$\langle\sigma_{\alpha}^{(1)}\otimes\sigma_{\beta}^{(2)}\rangle_{\rho}$.
It has been shown that only the expectation of $\sigma_{z}$ in the
phase qubit can be measured; however, the other direction
measurements can be achieved by applying a transformation that
maps the detected eigenvector onto the eigenvector of $\sigma_{z}$
before detecting. To obtain all 16 expectation values, nine
different settings have to be used.

A disadvantage of the tomography is that the operators required to
detect the entanglement are growing exponentially with the number of
qubits. However, if one knows about some priori information about
the generated entangled state, one can use entanglement witness
operator to distinguish and characterize the $N$-partite entangled
state\cite{Toth,Guhne,Chen}. The $W$ state and cluster states
proposed in the paper are actually two typical types of genuine
multipartite entangled states (i.e., the state whose reduced density
operator of any subsystem has rank larger than 1). The entanglement
witness $\mathcal{W}$ is an operator such that for every product
state
$$\text{Tr}(\rho\mathcal{W})\ge 0 \ \ \ \ (\rho\in
S_s)$$ with $S_s$ denoting the set of separable states. From the
definition of the operator $\mathcal{W}$, it is clear that the
witness has a positive or zero expectation value for all separable
states, and thus a negative expectation value signals the presence
of genuine multipartite entanglement. In order to measure the
witness $\mathcal{W}$ of the generating entanglement of the TLSs
proposed here, we should decompose the witness operator into a sum
of locally measurable operators. By appropriately constructing
$\mathcal{W}$, the required measurement settings are much less than
the requirement of the tomography.

As for the $N$-qubit W state denoted as $|W_N\rangle$, one of the
universal methods to construct the entanglement witness
$\mathcal{W}_{W_{N}}$  is given by
$$\mathcal{W}_{W_{N}}=\frac{N-1}{N}I-|W_{N}\rangle\langle W_{N}|$$
with $I$ denoting the identity operator. Especially, it has been
proven that the optimal decomposition of the witness
$\mathcal{W}_{W_{3}}$ is given by \cite{Guhne}
\begin{eqnarray}\label{}
\mathcal{W}_{W_{3}}=&&\frac{2}{3}I-|W_{3}\rangle\langle W_{3}|\nonumber\\
=&&\frac{1}{24}[17\cdot
I^{\otimes3}+7\cdot\sigma_{z}^{\otimes3}+3\cdot(\sigma_{z}II+I\sigma_{z}I+II\sigma_{z})\nonumber\\
&&+5\cdot(\sigma_{z}\sigma_{z}I+\sigma_{z}I\sigma_{z}+I\sigma_{z}\sigma_{z})\nonumber\\
&&-(I+\sigma_{z}+\sigma_{x})^{\otimes3}-(I+\sigma_{z}-\sigma_{x})^{\otimes3}\nonumber\\
&&-(I+\sigma_{z}+\sigma_{y})^{\otimes3}-(I+\sigma_{z}-\sigma_{y})^{\otimes3}].
\end{eqnarray}
This decomposition requires five measurement settings, namely
$\sigma_{z}^{\otimes3}$ and
$((\sigma_{z}+\sigma_{\eta})/\sqrt{2})^{\otimes3}$, $\eta=x,y$.
Since three qubits entangled states have not been realized in solid
state systems, the above optimal decomposition is very useful to
detect a purely genuine entangled state in near future experiments.
In addition, a universal method to construct the witness
$\mathcal{W}_{W_{N}}$ which requires $N^2-N+1$ measurement settings
is developed in Ref.\cite{Chen}.

We now turn to address a very efficient method to construct the
entanglement witness $\mathcal{W}_{C_{N}}$ of the $N$-qubit
cluster state $|C_N\rangle$. By using the stabilizing operators
$S_{j}^{(C_{N})}$ of the cluster state (i.e.,
$S_{j}^{(C_{N})}|C_{N}\rangle=|C_{N}\rangle$) defined as
\begin{eqnarray}
S_{1}^{(C_{N})}&=&\sigma_{x}^{(1)}\sigma_{z}^{(2)},\\
S_{j}^{(C_{N})}&=&\sigma_{z}^{(j-1)}\sigma_{x}^{(j)}\sigma_{z}^{(j+1)}
\ \
(j=2,3,\ldots,N-1),\\
S_{N}^{(C_{N})}&=&\sigma_{z}^{(N-1)}\sigma_{x}^{(N)},
\end{eqnarray}
one can construct the entanglement witness $\mathcal{W}_{C_{N}}$ ,
which detects genuine $N$-qubit entanglement around the $N$-qubit
cluster state, given by \cite{Toth}
\begin{equation}\label{W_c}
\mathcal{W}_{C_{N}}=3I-2\left[\prod_{\text{even}\
k}\frac{S_{k}^{(C_{N})}}{2}+\prod_{\text{odd}\
k}\frac{S_{k}^{(C_{N})}}{2}\right].
\end{equation}
A remarkable feature of this entanglement witness is that only two
local measurement settings, i.e.,
\begin{eqnarray*}
\sigma_x^{(1)} &\otimes& \sigma_z^{(2)}\cdots\sigma_x^{(j-1)}\otimes\sigma_z^{(j)}\cdots\sigma_x^{(N-1)}\otimes\sigma_z^{(N)},\\
\sigma_z^{(1)} &\otimes&
\sigma_x^{(2)}\cdots\sigma_z^{(j-1)}\otimes\sigma_x^{(j)}\cdots\sigma_z^{(N-1)}\otimes\sigma_x^{(N)},
\end{eqnarray*} are needed independent of the number of qubits.
Comparing with quantum state tomography that the number of
measuring settings increase exponentially with the number of
qubits, the required settings for the entanglement witness
increase at most polynomial with the number of qubits, and
specially for the cluster state, only two local measurement
settings are needed for any number of qubits.

\section{Generalization and Conclusion}

\begin{figure}[b]
\includegraphics[scale=0.45,width=7.5cm, angle=90]{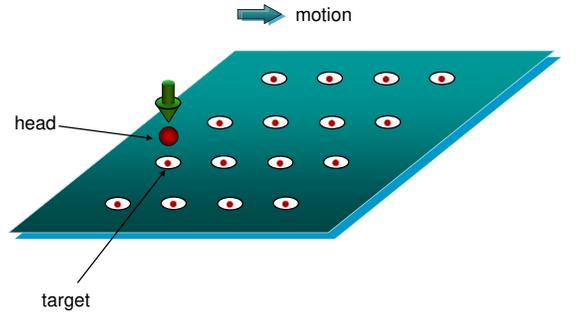}
\caption{Scalable quantum computer: two-dimensional array of
micro-traps.} \label{FIG.3}
\end{figure}

Actually the proposed approach to produce genuine multi-qubit
entanglement may be applied to a wide class of the candidates for
quantum computation, examples including trapped-ion quantum
computation and quantum computation based on $C_{60}$ etc.. For
concreteness, we generalize this method to achieve multi-qubit
entanglement of trapped ions proposed in Ref.\cite{Cirac}. Trapped
atomic ions remain one of the most attractive candidates for the
realization of a quantum computer, owing to their long-lived
internal qubit coherence. The central challenge now is to scale up
the number of trapped ion qubits. However, scaling the ion trap to
interesting numbers of ions poses significant difficulties.  In
Ref.\cite{Cirac}, the authors proposed an interesting scheme to
scale the ion qubits. As shown in Fig.3, ion qubits locate in a
two-dimensional array of micro-traps, where the distance between
the ions in the plane can be very large, since no direct
interaction between them is required. A different ion named the
head ion can move above the plane of the ion array. By switching
on a laser propagating in the perpendicular direction to the
plane, one can perform the two-qubit gate between the target ion
and the head ion. In particular, if the two-qubit iSWAP gate
described in the present paper can be performed between the head
ion and all ion qubits in the array, one can produce a large
number of the cluster state or $W$ state. IF the cluster state of
the ion qubits in this two-dimensional array can be achieved, an
one-way quantum computation may be implemented, since single qubit
gate for each ion qubits and the measurement operators can be
easily realized for trapped ions.

Furthermore, if each trapped ions in Fig.3 is replaced by a qubit
consisting of $C_{60}$\cite{Benjamin}, the genuine multi-qubit
entanglement for such qubits can also been achieved by using the
scheme we proposed here. Noted that it is difficult to realize the
strong couplings between all nearest neighbor qubits. So the
approach proposed here is promising.

In conclusion, we have presented an efficient scheme to realize
and detect the multi-qubit entangled states of TLSs locating
inside the Josephson phase qubit, and the proposed method can be
able to applied in a wide class of candidates for quantum
computation.

\section{Acknowledgements}
This work was supported by the NSFC (No. 10674049), the State Key
Program for Basic Research of China
(Nos. 2006CB921801 and 2007CB925204), and the RGC of Hong Kong (Nos.
HKU7051/06P and HKU7049/07P).

\end{document}